\RequirePackage{fix-cm}
\documentclass{svjour3}                     
\smartqed  
\usepackage{graphicx}
\newcommand{\be}{\begin{equation}}
\newcommand{\ee}{\end{equation}}
\def\tr{\mathop{\rm tr}\nolimits}

\def\dif{{\rm d}}

\begin{document}


\title{On the relativistic compressibility conditions}

\titlerunning{On the relativistic compressibility conditions}        

\author{Bartolom\'e Coll, Joan Josep Ferrando        \and
   Juan Antonio S\'aez 
}


\institute{Bartolom\'e Coll \at
              Departament d'Astronomia i Astrof\'{\i}sica, Universitat
de Val\`encia, \\ E-46100 Burjassot, Val\`encia, Spain. \and
Joan Josep Ferrando \at
              Departament d'Astronomia i Astrof\'{\i}sica, Universitat
de Val\`encia, \\ E-46100 Burjassot, Val\`encia, Spain. \\
Observatori Astron\`omic, Universitat
de Val\`encia, \\ E-46980 Paterna, Val\`encia, Spain. \\
                           \email{joan.ferrando@uv.es}           
           \and
          Juan Antonio S\'aez  \at
               Departament de Matem\`atiques per a l'Economia i l'Empresa,
Universitat de Val\`encia, \\E-46071 Val\`encia, Spain
}

\date{Received:  / Accepted: date}


\maketitle

\begin{abstract} 
The constraints imposed by the relativistic compressibility hypothesis on the square of the speed of sound in a medium are obtained. This result allows to obtain purely hydrodynamic conditions for the physical reality of a perfect energy tensor representing the energetic evolution of a perfect fluid in local thermal equilibrium. The results are applied to the paradigmatic case of the generic ideal gases. Then the physical reality of the ideal gas Stephani models is analyzed and the Rainich-like theory for ideal gas solutions is built.

\keywords{Thermodynamics - Compressibility conditions - Relativistic Perfect Fluids}
\end{abstract}

\section{Introduction}

In relativity, the gravitational field $g$ is related by Einstein equations to the energy content $T$ of the space-time. We know that not all energy tensors $T$ represent physically admissible energy contents, so that they have to be constrained by suitable {\em causal} and {\em energy conditions} which, in turn, restrict the physically admissible gravitational fields.

In addition, the energy content of particular physical media is described by specific energy tensors. For example,  a {\em perfect fluid}  imposes $T$ to be of the specific form:
\begin{equation}\label{perflud}
T = (\rho + p)u\otimes u + pg \, ,
\end{equation}
where $u$, $\rho$ and $p$ denote the {\em unit velocity} of the fluid, its {\em energy density} and its {\em pressure}, respectively. Obviously, such a particular choice of physical media additionally restricts the physically admissible gravitational fields. 

We are concerned here with thermodynamic perfect fluids in local thermal equilibrium. A {\em thermodynamic perfect fluid}  is a perfect fluid characterized, in addition to the above quantities, by its conserved {\em matter density}  $r$, {\em internal energy} $\epsilon$, {\em specific entropy} $s$ and {\em temperature} $\Theta$. And it is in {\em local thermal equilibrium}  (l.t.e.) if it verifies  the energy balance: 
\begin{equation}\label{balanceeq}
\Theta \dif s = \dif \epsilon + p \dif  v \, ,
\end{equation}
where $v= 1/r$ is the {\em specific volume}. It was shown a long time ago that, in order to be physically admissible, relativistic thermodynamic perfect fluids in l.t.e. have to verify the {\em relativistic compressibility conditions}:%
\footnote{See Sect. \ref{sec-cc} for details and nuances.} 
which may be written:
$$
\begin{array}{cc}
    {\rm H}_1:  & \qquad    (\tau'_p)_s < 0 \, , \qquad \quad (\tau''_p)_s > 0 \, , \qquad  \qquad \\ [9pt]
    {\rm H}_2:  &   (\tau'_s)_p > 0 \, ,   \qquad  
\end{array}
$$
where the {\em dynamic volume} $\tau = fv$ is the specific volume $v$ weighted by the {\em enthalpy index},\footnote{In this work the units are such that $c = 1$. We explicit it here in order to make clear that in the classical limit $c^2 \rightarrow \infty$, the dynamical volume $\tau$ reduces to the specific volume $v$.} 
$f= 1 + i/c^2$, the {\em specific enthalpy} $i$ being given by $i = \epsilon + pv$. 

In Newtonian gravitation, whatever the thermodynamic characterization of a perfect fluid may be, the Poisson equation states that the only physical quantity of a volume element of the fluid that causes or undergoes directly the gravitational field is its matter density. 

Correspondingly, in relativity, Einstein equations tell us that there is the particular combination of the three {\em hydrodynamic quantities} $u$, $\rho$ and $p$ in the {\em perfect energy tensor} $T$ given by (\ref{perflud})  which directly causes or undergoes the gravitational field. 

The above relativistic compressibility conditions involve thermodynamic, not only hydrodynamic, quantities, leading us to ask the following question: 

\begin{center}
{\em Do the relativistic compressibility conditions impose constraints \\ on the energy tensor $T$  of a thermodynamic  perfect fluid in l.t.e.?} 
\end{center}
Note that, via Einstein equations, this question is the equivalent of asking whether or not the relativistic compressibility conditions restrict the physically admissible gravitational fields.  

The analysis of this question is interesting for both, conceptual and practical reasons: 
%

- From a conceptual point of view, a relativistic thermodynamic perfect fluid may be considered as a {\em conservative and deterministic hydrodynamic perfect flow} endowed  with {\em subsidiary thermic quantities}.%
\footnote{See Sect. \ref{sec-t-flow} for details.} %
Thus, the relativistic compressibility conditions could impose restrictions on these subsidiary, non gravitational, quantities, on its hydrodynamic flow creating or undergoing the gravitational field or  on both ingredients. And it is important to determine which of these three possibilities is correct and to explicitly obtain the form of the corresponding constraints. This is the first objective of the present work.

- From a practical point of view, the solution to the {\em inverse problem} for perfect fluids (namely the determination of all the physical thermodynamic perfect fluids whose evolution in l.t.e. is described by a given perfect fluid energy tensor T) is tantamount to the acquisition  of all characteristic equations of state and thermodynamic quantities compatible with that given tensor $T$. It becomes crucial to know the degree of freedom or constraints to which these unknown thermodynamic relations and quantities are submitted. The second objective of the present work is to analyze and practically illustrate this situation. We shall do this for the particular case of ideal gases. 


In Sect. \ref{sec-t-flow} we introduce the concept of {\em hydrodynamic flow of a thermodynamic perfect fluid}, which gathers a significant result \cite{Coll-Ferrando-termo} \cite{fluperLTE}: the hydrodynamic quantities $\{u, \rho, p\}$ can be submitted to complementary hydrodynamic constraints that ensure the existence of the full set of thermodynamic quantities. In fact, these additional conditions state that the  {\em indicatrix of local thermal equilibrium}, $\chi \equiv \dot{p}/\dot{\rho}$, is a function of state, $\chi=\chi(\rho,p)$, and then it coincides with the square of the speed of sound. 

In Sect. \ref{sec-cc} we address the main goal of this paper, namely, the hydrodynamic flow approach to the relativistic compressibility conditions. Firstly we summarize the results by Israel \cite{Israel} and Lichnerowicz \cite{Lichnero-1}, and then we analyze separately two cases: the intrinsic barotropic media and the non barotropic ones. We show that the compressibility conditions H$_1$ can be stated in terms of the indicatrix function $\chi(\rho,p)$, that is, they impose constraints on the hydrodynamic flow. Nevertheless, condition H$_2$ generically only imposes constraints on the thermodynamic subsidiary quantities.

In Sect. \ref{sec-cc-idealgas} we study in detail the specific case of a generic ideal gas. We start from our hydrodynamic characterization of an ideal gas \cite{fluperLTE}, which requires that the indicatrix function $\chi \equiv \dot{p}/\dot{\rho}$ depends on the hydrodynamic variable $\pi=p/\rho$, $\chi= \chi(\pi)$. Then, the compressibility condition H$_2$ on the ideal gas thermodynamic quantities can also be stated in terms of the function $\chi(\pi)$.

In Sect. \ref{sec-Rainich} the results in the previous section enable us to acquire the Rainich approach for the perfect fluid solutions of the Einstein equations performing an ideal gas which fulfills the relativistic compressibility conditions.

In Sect. \ref{sec-idealgas-Stephani} we apply our results to analyze the physical reality of the ideal gas Stephani models obtained in \cite{C-F}. In addition to the full compressibility conditions, we impose additional constraints ensuring a good performance at high and low temperatures.

Finally, in Sect. \ref{sec-remarks} we comment on the relevance of the results obtained here.

In this paper we work on an oriented spacetime with a metric tensor
$g$ of signature $\{-,+,+,+\}$. We denote with the same symbol any tensor and its associated ones by raising and lowering indexes with the metric tensor. For the metric product of two vectors, we write
$(x,y) = g(x,y)$, and we put $x^2 = g(x,x)$. If $S$ is a 2-tensor, $S(x)$ denotes the vector with covariant components $S_{\alpha \beta} x^{\beta}$, $S(x,y) = S_{\alpha \beta} x^{\alpha}y^{\beta}$, and $S^2$ the 2-tensor with components $(S^2)_{\alpha \beta} = S_{\alpha}^{\ \lambda} S_{\lambda \beta}$.


%
\section{Hydrodynamic flow of a thermodynamic perfect fluid in l.t.e.}
\label{sec-t-flow}
%

The divergence-free condition for the {\em perfect energy tensor}\, $T$ given by (\ref{perflud}),
\begin{equation}\label{hydroenergyconserv}
 \nabla \cdot ( (\rho + p)u\otimes u + pg) = 0\, ,
\end{equation}
 leads to a conservative system of four equations for the five hydrodynamic quantities $\{u,\rho,p\}$. 
For thermodynamic perfect fluids in l.t.e. the {\em deterministic closure}%
\footnote{That is to say, the set of complementary equations to be added to the system in order that it admits unicity of the Cauchy problem.} %
to this system   is obtained \cite{Eckart} by decomposing the energy density $\rho$ in terms of the matter density $r$ and the specific internal energy $\epsilon$, $\rho= r(1+\epsilon)$, and relating them by the balance equation (\ref{balanceeq}), which can be written as:
\begin{equation}
\Theta \dif s = (1/r) \dif \rho + (\rho+p) \dif(1/r) \, .   \label{re-termo}
\end{equation}
As a consequence, only two thermodynamic quantities are independent and they define a thermodynamic plane. Equations (\ref{hydroenergyconserv}) and (\ref{re-termo}) imply the equivalence between matter conservation and local adiabatic evolution, which can be stated, respectively, by the following conditions:
\begin{equation}\label{determeqs}
\dot{r} + r \theta = 0 \, ,   \qquad \qquad  \dot{s} = 0 \, ,  \label{c-massa-s}
\end{equation}
$\theta = \nabla_\alpha u^\alpha$ being the expansion of the fluid, and where a dot denotes the directional derivative, with respect to $u$, of a quantity $q$, $\dot{q} = u(q) = u^{\alpha} \partial_{\alpha} q$. 
When (\ref{re-termo}) holds, any of these two equations may play the role of deterministic closure to the conservation equations (\ref{hydroenergyconserv}), and the whole set  constitutes the deterministic {\em fundamental system of the relativistic hydrodynamics}. 

As any of the deterministic closures (\ref{determeqs}) requires  (\ref{re-termo}),  they generate  the impression that any l.t.e. closure to the conservation equations  (\ref{hydroenergyconserv}) necessarily needs the introduction of  new, {\em thermodynamic}, quantities. Such an impression is incorrect, as already shown by a previous result \cite{Coll-Ferrando-termo} (see also the recent paper \cite{fluperLTE}):

\begin{theorem}\label{ColFer1989}
The necessary and sufficient condition for a conservative perfect energy tensor $T = (\rho + p)u\otimes u + pg$ to describe a thermodynamic perfect fluid in l.t.e. is that its hydrodynamic quantities $\{u,\rho,p\}$ fulfill the {\em hydrodynamic sonic condition}:
\begin{equation}\label{soundeq}
(\dot{\rho} \dif \dot{p} - \dot{p} \dif \dot{\rho}) \wedge \dif\rho 
\wedge \dif p = 0       \, .    \label{h-lte}
\end{equation}
\end{theorem}

The hydrodynamic sonic condition (\ref{soundeq}) is a deterministic closure for the conservative system (\ref{hydroenergyconserv}) so that {\em the evolution of the thermodynamic perfect fluid in l.t.e. is uniquely determined by the differential system ${\cal{H}} \equiv \{(\ref{hydroenergyconserv}),(\ref{soundeq})\}$ in the hydrodynamic quantities $\{u,\rho,p\}$}. We shall call this system ${\cal{H}}$ the {\em hydrodynamic flow} of the thermodynamic perfect fluid in l.t.e.

Theorem \ref{ColFer1989} states that the solutions of the system ${\cal{H}}$  characterize biunivocally {\em the evolution} of the thermodynamic perfect fluid in l.t.e. But they do not characterize biunivocally the thermodynamic perfect fluid in l.t.e. {\em itself},  because they may also be verified by other, different, continuous media. 

As follows from the above considerations, {\em a thermodynamic perfect fluid in l.t.e. is biunivocally determined by its hydrodynamic flow ${\cal{H}}$ endowed with two quantities $r$ and $s$ depending on the two hydrodynamic variables $\rho$ and $p$ and submitted to the equations} (\ref{determeqs}).

It is important to observe that, the {\em hydrodynamic sonic condition} (\ref{soundeq}) being included in the hydrodynamic flow ${\cal{H}}$ as a deterministic closure, {\em the hydrodnamic quantities determine completely the evolution of the fluid} meanwhile  the thermodynamic quantities appear as  subsidiary variables specifically allowing the characterization of the sole {\em thermodynamic properties} of the fluid. In this sense, the equations (\ref{determeqs})  no longer play the role of deterministic closures for the conservative system (\ref{hydroenergyconserv}) but are ``reduced'' to simple {\em evolution equations} for the subsidiary thermodynamic quantities $r$ and $s$.

The interest of the hydrodynamic flow approach to the l.t.e. of a thermodynamic perfect fluid has been widely pointed out in \cite{fluperLTE} and some applications have recently been developed \cite{CFS-CIG} \cite{CFS-PSS}. From this approach, the {\em indicatrix of local thermal equilibrium} \cite{fluperLTE}, $\chi \equiv \dot{p}/\dot{\rho}$, plays a central role. We remember here that, when $\dot{\rho} \not= 0$, $\chi$ is a function of state, $\chi=\chi(\rho,p)$, and that it then coincides with the square of the speed of sound and the hydrodynamic l.t.e. characterization (\ref{soundeq}) identically holds; for this reason we also call this characterization the hydrodynamic sonic condition.


\section{Relativistic compressibility conditions}
\label{sec-cc}


In his study of shock waves in a relativistic thermodynamic perfect fluid in l.t.e., Israel \cite{Israel} was lead to impose the above relativistic compressibility conditions
${\rm H_1}$ and ${\rm H_2}$ in order to obtain a coherent theory.%
\footnote{Namely, one in which (i) shock velocities are always less than the light constant $c$, (ii) shocks are compressive supersonic waves with increasing entropy across them, and (iii) the state after the shock is univocally determined by the state before the shock.} %
Later, Lichnerowicz \cite{Lichnero-1} showed that, in {\em relativistic magnetohydrodynamics}, these relativistic compressibility conditions lead to analogous results.

The relativistic compressibility conditions ${\rm H_1}$ and ${\rm H_2}$ in terms of the dynamic specific volume $\tau$ have the same form that the Weyl classical ones \cite{Weyl}, $(v'_p)_s < 0$, $(v''_p)_s > 0$,  $(v'_s)_p > 0$, in terms of the specific volume $v$ and, of course, both conditions coincide in the classical limit $c^2 \rightarrow \infty$ (see footnote 2). These facts suggest that the dynamic specific volume $\tau$ is a leading variable in relativistic thermodynamics.

Following this idea, the relativistic compressibility conditions ${\rm H_1}$ and ${\rm H_2}$, and a suitable relativistic definition of exothermic reaction in terms of the dynamical volume $\tau$, allowed one of us, Coll \cite{coll-T} \cite{coll-HP}, to construct the relativistic theory of deflagrations and detonations in a magnetohydrodynamic fluid.\footnote{An abridged English version may be found in \cite{Anile}, Ch. 8, \S 3.}

From Synge's results \cite{Synge} on the relativistic Boltzmann gas, it follows that this gas verifies the relativistic compressibility conditions ${\rm H_1}$ and ${\rm H_2}$. Israel \cite{IsraelTh} showed that the first inequality of the compressibility conditions ${\rm H_1}$  remains true also for relativistic Bose and Fermi gases. And Lucquiaud \cite{Lucquiaud1} \cite{Lucquiaud2}  proved that both relativistic compressibility conditions ${\rm H_1}$ and ${\rm H_2}$ are equally satisfied by relativistic Boltzmann, Bose and Fermi gases.

In \cite{Thorne}, Thorne extended (for vanishing magnetic field) the work of Israel and Lichnerowicz, showing that their essential results could be obtained  in absence of the compressibility condition ${\rm H_2}$. Subsequently, Lichnerowicz \cite{LichnerowiczJMP} \cite{Lichnero-2} reworked the relativistic theory of shock waves in magnetohydrodynamics under both hypothesis,
${\rm H_2}$ and it inverse, that we denote  ${\rm \bar{H}_2}$:
$$
        {\rm \bar{H}_2:}  \qquad   \qquad  (\tau'_s)_p < 0 \,\,  .   \qquad \qquad 
$$

For any medium in l.t.e. one has\footnote{This expression follows from the integrability condition of (\ref{balanceeq}), $(v'_s)_p = (\Theta'_p)_s$, and the relation $(v'_s)_p = (v'_\tau)_p (\tau'_s)_p$ deduced from $v= v(\tau(s,p),p)$.} $(v'_\tau)_p = (\Theta'_p)_s/(\tau'_s)_p$.
Consequently, if the medium has a classical limit, $\tau \rightarrow v$ and $(v'_{\tau})_p \rightarrow 1$, and then we have, necessarily, $(\Theta'_p)_s < 0$. Thus, for a medium that holds ${\rm \bar{H}_2}$, any isentropic evolution implies that its temperature decreases when pressure increases. In the scientific literature such media seem scarce. Anyway, their possible existence leads us, in what follows, to take into account also this condition ${\rm \bar{H}_2}$.

We want now to analyze to what extend the relativistic compressibility conditions ${\rm H_1}$ and  ${\rm H_2}$ or ${\rm \bar{H}_2}$ concern the hydrodynamic flow of the perfect fluid in l.t.e. or only its subsidiary thermodynamic quantities. For this task, we must consider separately the barotropic and non barotropic cases.


\subsection{Compressibility conditions for barotropic media}
\label{sec-cc-bar}

Let us consider a barotropic fluid submitted to the barotropic equation of state $p = \phi(\rho)$.\footnote{We avoid the case $\rho=\rho_0$, which leads to an unphysical thermodynamic scheme since it does not meet any of the above relativistic compressibility conditions.}  In this case the characteristic equation $r=r(\rho,s)$ takes the expression \cite{fluperLTE}:
\begin{equation}
r(\rho,s) = \frac{G(\rho)}{R(s)}, \qquad \quad	 G(\rho) = \exp \left[\int \! \!\frac{d \rho}{\rho+ \phi(\rho)}\right]  \, .
\label{r-ro-s-bar}
\end{equation}
The dynamic specific volume $\tau$ may be written
\begin{equation} \label{dyn-vol}
\tau = \frac{1}{r^2}(\rho+p) \, ,
\end{equation}
and by derivation taking into account (\ref{r-ro-s-bar}) we obtain:
\begin{eqnarray}
(\tau'_p)_s = \frac{1}{r^2\phi'}(\phi'- 1) \, ,
 \label{cc-1a-bar} \\[1mm]
(\tau''_p)_s = \frac{1}{r^2(\rho+p)\phi'^3} [\phi''(\rho+p)+ 2 \phi' (1- \phi')] \, ,
 \label{cc-1b-bar} \\[1mm]
 \qquad (\tau'_s)_p = \frac{2 \tau}{R(s)} R'(s) \, ,
\label{cc-2-bar}
\end{eqnarray}
and we have:
\begin{theorem} \label{theo-cc-bar}
For a barotropic perfect fluid in l.t.e. with equation of state $p = \phi(\rho)$, the relativistic compressibility conditions ${\rm H_1}$ and ${\rm H_2}$ or ${\rm \bar{H}_2}$ may be written as:
\begin{equation}\label{cc-1-bar}
{\rm H_1}: \quad 0 < \phi' < 1 \, , \quad   (\rho+p)\phi''+ 2 \phi' (1- \phi') > 0
\end{equation}
and
\begin{equation}\label{cc-2-bar}
{\rm H_2}: \quad R'(s) >0  \qquad \ \mbox{ or } \qquad \  {\rm \bar{H}_2}: \quad R'(s) <0 \, ,  
\end{equation}
where $R(s)$ is the entropy denominator of the characteristic equation $r = r(\rho,s)$ given by {\rm(\ref{r-ro-s-bar})}.
\end{theorem}

As a consequence of this theorem and Theorem \ref{ColFer1989}, one has:
\begin{theorem} \label{theo-cc-bar-b}
The necessary and sufficient condition for a barotropic perfect energy tensor $T$ to represent the evolution in l.t.e. of a barotropic fluid verifying ${\rm H_1}$ and ${\rm H_2}$ or ${\rm \bar{H}_2}$ is ${\rm H_1}$. Any such energy tensor $T$ may be endowed with a characteristic equation $r = r(\rho,s)$ verifying ${\rm H_2}$ or ${\rm \bar{H}_2}$, equally.
\end{theorem}

Before studying the non barotropic case, let us note three points:
\begin{itemize}
\item[{\bf i)}]
The expression (\ref{cc-1-bar}) of the conditions H$_1$ is known and can be found in the literature (see for example \cite{Anile}). Note that $\phi'$ is the square of the speed of sound, $c_s^2 = (p'_{\rho})_s = \phi'(\rho)$.
\item[{\bf ii)}]
Conditions (\ref{cc-1-bar}) exclusively involve hydrodynamic quantities. Thus, the relativistic compressibility conditions ${\rm H_1}$ only concern the hydrodynamic flow of the perfect fluid in l.t.e.  Nevertheless, constraint H$_2$ (or ${\rm \bar{H}_2}$) imposes an additional condition on its subsidiary thermodynamic quantities and it does not restrict the perfect energy tensor.
\item[{\bf iii)}] It is worth remarking that Theorem \ref{theo-cc-bar} only applies for {\em  intrinsic} barotropic fluids, that is, for barotropic perfect energy tensors where $p=\phi(\rho)$ is considered to be an equation of state.%
    \footnote{And not an evolution path in the thermodynamic plane of a more general equation of state.} Otherwise, a barotropic relation $p=\phi(\rho)$ that does not fulfill the expressions (\ref{cc-1-bar}) could be an admissible evolution path of a non barotropic fluid which fulfills the (general) compressibility conditions H$_1$. Conversely, a non barotropic perfect fluid which does not fulfill the compressibility hypothesis H$_1$ could admit a barotropic evolution that fulfills conditions (\ref{cc-1-bar}). We will return to this fact at the end of the following subsection.
\end{itemize}
%


\subsection{Compressibility conditions for non barotropic media}
\label{sec-cc-nonbar}

Under the non barotropic assumption, we can consider the thermodynamic quantities $(\rho,p)$ as coordinates in the thermodynamic plane.\footnote{For a function of state $z(\rho,p)$ we write $z_{\rho}' = (z_{\rho}')_p$, $z_p' = (z_p')_{\rho}$.} As we have seen, a thermodynamics is determined by two functions $r(\rho,p)$ and $s(\rho, p)$ that are solution to equations (\ref{c-massa-s}) and are constrained by the thermodynamic relation (\ref{re-termo}). If one of these functions of state is known, the square of the speed of sound can be obtained as \cite{fluperLTE}:
\begin{equation}
c_s^2 = \chi(\rho,p) \equiv - \frac{s_{\rho}'}{s_p'}= \frac{1}{r_p'}\left[\frac{r}{\rho+p} - r_{\rho}' \right] \, .  \label{chi-s-r}
\end{equation}
And conversely, it is worth remarking the simplicity with which the square of the speed of sound  $\chi(\rho,p)$ allows to control the  different thermodynamic schemes with which a hydrodynamic flow may be endowed. Indeed, as seen in \cite{fluperLTE}, the mass density $r= r(\rho,p)$ and the entropy $s=s(\rho,p)$ are of the form $r=\bar{r}R(\bar{s})$ and $s=s(\bar{s})$, where $\bar{r}(\rho,p)$ is any particular solution of equation (\ref{chi-s-r}) in $r$, and $R(\bar{s})$ and $s(\bar{s})$ are arbitrary functions of any particular solution $\bar{s}(\rho,p)$ of the  equation (\ref{chi-s-r}) in $s$.

From (\ref{chi-s-r}) we obtain that, for any function of state $z=z(s,p) = z(\rho,p)$, we have:
\begin{equation}
(z_{s}')_p =  \frac{z_{\rho}'}{s_{\rho}'} \, , \qquad (z_{p}')_s =  z_p' - z_{\rho}' \frac{s_{p}'}{s_{\rho}'} =  z_p' + \frac{1}{\chi} z_{\rho}'  \, .  \label{z}
\end{equation}
Moreover, from (\ref{dyn-vol}) we obtain:
\begin{equation}
\tau_{\rho}' =  -\frac{2}{r^3} r_{\rho}' (\rho+p) + \frac{1}{r^2}  \, , \qquad \tau_p' =  -\frac{2}{r^3} r_p' (\rho+p) + \frac{1}{r^2}  \, .  \label{tau-ro-p}
\end{equation}
Then, from (\ref{z}) (with $z=\tau$) and (\ref{tau-ro-p}) we have:
\begin{equation}
(\tau_{p}')_s = \xi(\rho, p) \equiv \frac{1}{r^2\chi}(\chi - 1)   \, .  \label{cc-1a-nobar}
\end{equation}

From the function of state $\xi(\rho, p)$ given above we derive:
\begin{equation}
\xi_{\rho}' =  \frac{1}{r^2 \chi^2} \left[\chi_{\rho}' - \frac{2}{r} \chi(\chi-1)  r_{\rho}' \right]  \, , \qquad \xi_{p}' =  \frac{1}{r^2 \chi^2} \left[\chi_{p}' - \frac{2}{r} \chi(\chi-1)  r_{p}' \right]  \, .  \label{xi-ro-p}
\end{equation}
Then, making use of (\ref{z}) (with $z=\xi$), (\ref{tau-ro-p}) and (\ref{chi-s-r}), we have:
\begin{equation}
(\tau_{p}'')_s =  \frac{1}{r^2(\rho+p)\chi^3}\left[(\rho+p)(\chi \chi_{p}' + \chi_{\rho}') + 2 \chi(1-\chi) \right]   \, .  \label{cc-1b-nobar}
\end{equation}

On the other hand, from (\ref{z}) (with $z=\tau$) and (\ref{tau-ro-p}) we obtain:
\begin{equation}
(\tau_{s}')_p =  \frac{1}{r^2 s_{\rho}'}\left[1 - \frac{2(\rho+p)}{r} r_{\rho}' \right]   \, .  \label{cc-2-nobar}
\end{equation}
Consequently, if we take into account expressions (\ref{cc-1a-nobar}), (\ref{cc-1b-nobar}) and (\ref{cc-2-nobar}) we arrive to the following result:

\begin{theorem} \label{theo-cc-nobar}
For a non barotropic perfect fluid in l.t.e., the relativistic compressibility conditions
${\rm H_1}$ and ${\rm H_2}$ or ${\rm \bar{H}_2}$ may be written as:
\begin{equation}
{\rm H_1}: \quad
0 < \chi < 1 \, , \qquad   (\rho+p)(\chi \chi_{p}' + \chi_{\rho}') + 2 \chi(1-\chi) > 0   \, ,       \label{cc-1-nonbar}
\end{equation}
and
\begin{equation}\label{cc-2-nonbar}
{\rm H_2}: \qquad I >0  \qquad \quad \mbox{ or } \qquad \quad   {\rm \bar{H}_2}: \qquad \ I
 <0 \, ,  \qquad \qquad \
\end{equation}
where $\chi$ is the square of the speed of sound as a function of state depending on $(\rho,p)$, $\chi=\chi(\rho,p)$, and $I = I(\rho,p)$ is the function that depends on $r=r(\rho,p)$ and  $s=s(\rho,p)$ as:
\begin{equation}
I(\rho,p)  \equiv  \frac{1}{s_{\rho}'}[r - 2(\rho+p)r_{\rho}']\, .  \label{cc-2b-nonbar}
\end{equation}
\end{theorem}
If we take into account the expression of the temperature $\Theta(\rho,p)$ in terms of the functions $r(\rho,p)$ and $s(\rho,p)$ \cite{fluperLTE}, the function of state $I(\rho,p)$ given in (\ref{cc-2b-nonbar}) can be written as $I = r[2r \Theta - 1/s_{\rho}']$. Then, we can obtain an equivalent expression for the condition H$_2$ which can be useful in subsequent applications \cite{CFS-PSS}: 
\begin{corollary} \label{cor-H2-Theta}
The compressibility conditions ${\rm H}_2$ or $\bar{{\rm H}}_2$ can be written as:
\begin{equation}\label{cc-2-theta}
{\rm H_2}: \qquad 2 r \Theta > \frac{1}{ s_{\rho}'}  \qquad \quad \mbox{ or } \qquad \quad   {\rm \bar{H}_2}: \qquad \  2 r \Theta < \frac{1}{s_{\rho}'} \, .  \qquad \qquad \
\end{equation}
\end{corollary}
Note that, under the physical requirements $\Theta >0$ and $r>0$, a sufficient condition for ${\rm H}_2$ is $s_{\rho}' < 0$.

We know that two different pairs $(\bar{r},\bar{s})$ and $(r,s)$, which are solution of equations (\ref{chi-s-r}), are related by $r=\bar{r}R(\bar{s})$ and $s=s(\bar{s})$. Then, the corresponding functions of state $\bar{I} \equiv \bar{I}(\rho,p)$, and $I \equiv I(\rho,p)$ given in (\ref{cc-2b-nonbar}) are related by:
\begin{equation}
I = \frac{1}{s'({\bar{s}})}[R(\bar{s}) \bar{I}  - 2\bar{r}(\rho+p)R'(\bar{s})]    \, .  \label{cc2-I}
\end{equation}
This expression implies that we can take the arbitrary function $R(\bar{s})$ such that $I$ does not vanish, and thus either ${\rm H_2}$ or ${\rm \bar{H}_2}$ is fulfilled. Then, any  of them may be selected by choosing the arbitrary function $s=s(\bar{s})$ such that $s'(\bar{s})$ has the appropriate sign.  Consequently, we have:
\begin{lemma} \label{lemma-H1-H2}
Any function of state $\chi(\rho,p)$ admits thermodynamic schemes verifying any of the relativistic compressibility conditions ${\rm H_2}$ or ${\rm \bar{H}_2}$.
\end{lemma}

Let us consider a non barotropic ($d\rho \wedge d p \not=0$) and non isoenergetic ($\dot{\rho} \not= 0$) energy tensor $T$. Then, we can define the {\em indicatrix function} $\chi \equiv \dot{p}/\dot{\rho}$. When $\chi$ is a function of state ($d \chi \wedge d\rho \wedge d p =0$), Theorem \ref{ColFer1989} implies that  the hydrodynamic sonic condition holds, and then $\chi$ coincides with the square of the speed of sound, $\chi = \chi(\rho,p)$, of the fluids for which $T$ is a possible evolution. Theorem \ref{theo-cc-nobar} and Lemma \ref{lemma-H1-H2} apply for these thermodynamics and, taking into account the results in \cite{fluperLTE}, we have:
\begin{theorem} \label{theo-cc-nonbar-chi}
The necessary and sufficient condition for a non barotropic  and non isoenergetic perfect  energy tensor $T$ to represent the evolution in l.t.e. of a perfect fluid verifying the relativistic compressibility conditions ${\rm H_1}$ and ${\rm H_2}$ or ${\rm \bar{H}_2}$ is that the indicatrix function $\chi \equiv \dot{p}/\dot{\rho}$ be a function of state, $\chi = \chi(\rho,p)$, verifying expression {\em(\ref{cc-1-nonbar})} of the relativistic compressibility conditions ${\rm H_1}$.

 Any such energy tensor $T$ may be endowed with thermodynamic quantities $r=r(\rho,p)$ and  $s=s(\rho,p)$ solutions of {\em (\ref{chi-s-r})} such that the relativistic compressibility conditions ${\rm H_2}$ or ${\rm \bar{H}_2}$ hold equally.
\end{theorem}

When $\dot{\rho}=0$ and $\dot{p}\not=0$ we have necessarily \cite{fluperLTE} $r=r(\rho)$ and $s=s(\rho)$, and then the compressibility conditions H$_1$ do not hold. If we have an isobaroenergetic evolution, $\dot{\rho}=\dot{p}=0$, the associated thermodynamic schemes are defined by an arbitrary matter density $r=r(\rho,p)$ and a specific entropy $s=s(\rho,p)$ solution to the last equality in (\ref{chi-s-r}) \cite{fluperLTE}. But, alternatively, we can give an arbitrary speed of sound, namely its square $\chi = \chi(\rho,p)$, which constraints $r=r(\rho,p)$ and $s=s(\rho,p)$ by means of (\ref{chi-s-r}). Then, the results in \cite{fluperLTE} and Theorem \ref{theo-cc-nobar} and Lemma \ref{lemma-H1-H2} imply:

\begin{theorem} \label{theo-cc-nonbar-iso}
A non barotropic and isoenergetic perfect energy tensor $T$ represents the evolution in l.t.e. of a perfect fluid submitted to the relativistic compressibility conditions ${\rm H_1}$ and ${\rm H_2}$ or ${\rm \bar{H}_2}$ if, and only if, it is isobaroenergetic, $\dot{\rho}=\dot{p}=0$.

For it, the admissible general thermodynamic quantities $r=r(\rho,p)$ and  $s=s(\rho,p)$ are those for which {\em(\ref{chi-s-r})} defines a square of the speed of sound  $\chi = \chi(\rho,p)$ verifying expression {\em(\ref{cc-1-nonbar})} of the relativistic compressibility condition ${\rm H_1}$. Then, {\em(\ref{cc2-I})} may generate new quantities $r=r(\rho,p)$ and  $s=s(\rho,p)$ verifying ${\rm H_2}$ or ${\rm \bar{H}_2}$, equally.
\end{theorem}

We finish this section with some suitable comments on the above results :
\begin{itemize}
\item[{\bf i)}]
Note that Theorems \ref{theo-cc-nonbar-chi} and \ref{theo-cc-nonbar-iso} provide a purely hydrodynamic characterization of the perfect energy tensors $T$ which are the evolution in l.t.e. of a thermodynamic perfect fluid submitted to the compressibility conditions ${\rm H_1}$ and ${\rm H_2}$ or ${\rm \bar{H}_2}$.
In other words, Theorems \ref{theo-cc-nonbar-chi} and \ref{theo-cc-nonbar-iso} characterize all the hydrodynamic flows able to correspond to thermodynamic perfect fluids submitted to the relativistic compressibility conditions.
This characterization imposes conditions (\ref{cc-1-nonbar}) on the square of the speed of sound $\chi(\rho,p)$.
In the non isoenergetic case, this function is determined by the hydrodynamic
 flow ${\cal{H}} \equiv \{(\ref{hydroenergyconserv}),(\ref{soundeq})\}$ through the indicatrix function $\chi(\rho,p)=\chi \equiv \dot{p}/\dot{\rho}$.
In the isoenergetic case, $\dot{\rho}=0$, we have necessarily $\dot{p}=0$, the indicatrix function is indeterminate, and the compressibility conditions does not impose any condition on the hydrodynamic flow ${\cal{H}}$. The thermodynamics associated with any of these isobaroenergetic flows are constrained by conditions (\ref{cc-1-nonbar}) that now define the square of the speed of sound $\chi(\rho,p)$.

\item[{\bf ii)}]
In the inverse problem \cite{fluperLTE} under consideration, ${\rm H_1}$ imposes the constraint (\ref{cc-1-nonbar}) on the function $\chi(\rho,p)$ and ${\rm H_2}$ or ${\rm \bar{H}_2}$ imposes the constraint (\ref{cc-2-nonbar}-\ref{cc-2b-nonbar}) to the, non hydrodynamic, thermodynamic  quantities $r$ and $s$. And, as stated in Lemma \ref{lemma-H1-H2}, for {\em any} given hydrodynamic function $\chi(\rho,p)$, there {\em always} exist thermodynamic functions $r$ and $s$ that fulfill ${\rm H_2}$ or ${\rm \bar{H}_2}$. 
Nevertheless, in {\em restricted} inverse problems \cite{fluperLTE} for specific families of perfect fluids, characterized by a relation involving  hydrodynamic as well as thermodynamic quantities, the compressibility conditions  ${\rm H_2}$ or ${\rm \bar{H}_2}$ may restrict strongly the hydrodynamic function $\chi \equiv \dot{p}/\dot{\rho}$ (this is the case, for example, of  the ideal gas. See next section).
\item[{\bf iii)}]
From (\ref{cc2-I}), the thermodynamic fluids that do not fulfill neither  ${\rm H_2}$ nor ${\rm \bar{H}_2}$ are those verifying $I = 0$, where $I$ is given in (\ref{cc-2b-nonbar}). It follows from this expression of $I$ that the matter density is of the form $r=r(\rho,p) = f ( p ) (\rho+p)^{1/2}$, where $f(p )$ is any arbitrary positive function. Then, from (\ref{chi-s-r}), the square of the speed of sound is given by $\chi(\rho,p) = [1+(\rho+p)F ( p )]^{-1}$, with $F \equiv 2 f'( p )/f( p)$, and it verifies ${\rm H_1}$ if and only if $F( p )>0$ and $F'( p ) < F^2(p)$. In fact, from (\ref{dyn-vol}), the function $f(p)$ and  the dynamic volume $\tau$ are related by $\tau = f^{-2}(p)$, which allows to directly see that $(\tau'_s)_p = 0$ i.e. that neither  ${\rm H_2}$ nor ${\rm \bar{H}_2}$ are verified.
\item[{\bf iv)}] For a non barotropic fluid, a barotropic evolution can be defined by a relation $s=s(\rho)$. Then, if $p=p(\rho,s)$, the barotropic relation between $\rho$ and $p$ is given by $p = \phi(\rho) \equiv p(\rho,s(\rho))$. We have, $c_s^2 = (p'_{\rho})_s \not= \phi'(\rho)$. Thus, $\phi'(\rho)$ is not the square of the speed of sound in this case. Consequently, Theorem \ref{theo-cc-bar} concerns barotropic energy tensors that
describe the evolution of intrinsic barotropic fluids, and Theorem \ref{theo-cc-nobar} concerns non barotropic energy tensors describing the evolution of non barotropic fluids. The cases of barotropic energy tensors that describe  barotropic evolutions of non barotropic fluids require  specific individual studies. Note that Theorem \ref{theo-cc-nobar} remains valid in this case and it gives restrictions on the square of the speed of sound. 
\item[{\bf v)}]
We know \cite{fluperLTE} that every barotropic and isobaroenergetic ($\dot{\rho}=\dot{p}=0$) perfect energy tensor represents the evolution in l.t.e. of {\em any} perfect fluid. Consequently, perfect fluids verifying the constraints (\ref{cc-1-nonbar}) and (\ref{cc-2-nonbar}-\ref{cc-2b-nonbar}) could have a particular barotropic evolution of the form  $p=\phi(\rho)$ that is not subjected to constraints (\ref{cc-1-bar}). For example, this is the case of any physical fluid evolving at constant pressure, $\phi(\rho)=p_0$. On the other hand, non barotropic fluids failling to verify ${\rm H_1}$, i.e. that do not  fulfill the constraints (\ref{cc-1-nonbar}), might have a barotropic evolution  $p=\phi(\rho$) satisfying (\ref{cc-1-bar}) (see at the end of next section for a specific example).
\end{itemize}


\section{Compressibility conditions for a generic ideal gas}
\label{sec-cc-idealgas}

A {\em generic ideal gas}%
\footnote{The equation of state (\ref{gas-ideal}) not being a {\em characteristic} equation of state in the Gibbs' sense, is insufficient to define completely the thermodynamics of the fluid (it is usually completed with a particular expression of the internal energy $\epsilon$  as a function of the temperature $\Theta$). It does not define {\em an} ideal gas, but anyone of the whole class. To reflect this fact, we call it a {\em generic ideal gas}.} %
 is described by the equation of state:
\begin{equation}
p = kr\Theta  \, , \qquad \quad    k \equiv {k_B \over m} \,  .  \label{gas-ideal}
\end{equation}
In \cite{fluperLTE} we have studied the expression of the speed of sound and the other thermodynamic quantities of a generic ideal gas in terms of the hydrodynamic quantities $(\rho, p)$. We present in the next lemma a slightly different version of these results, which will be useful in order to analyze the relativistic compressibility conditions.
\begin{lemma}  \label{lemma-ideal}
The square of the speed of sound $\chi$ of a non barotropic generic ideal gas depends on the sole hydrodynamic variable $\pi \equiv p/\rho$, $\chi=\chi(\pi) \not= \pi$. Then, in terms of the hydrodynamic quantities $(\rho,p)$, the specific internal energy
$\epsilon$, the temperature $\Theta$, the matter density $r$ and the specific entropy $s$ are given, respectively, by:
\begin{eqnarray}
\epsilon(\rho,p) = \epsilon(\pi) \equiv e(\pi)-1 \, , \qquad \quad
\Theta(\rho,p) = \Theta(\pi) \equiv {\pi \over k} e(\pi) \, , \label{e-t-ideal} \\
r(\rho,p) = {\rho \over e(\pi)} \, ,  \qquad  \qquad    \qquad  \
\quad s(\rho,p) = k \ln \frac{f(\pi)}{\rho} \, , \quad  \quad \ \  \label{r-s-ideal}
\end{eqnarray}
the generating functions $e(\pi)$ and $\phi(\pi)$ being, respectively,
\begin{eqnarray}
e(\pi) = e_0 \exp\{\! \! \int \! \! \psi(\pi)d\pi \} \, , \qquad  \qquad \psi
(\pi) \equiv \frac{\pi}{(\chi(\pi)-\pi)(\pi+1)} \, \ ; \label{e-pi} \\
f(\pi) = f_0 \exp\{\! \! \int \! \! \phi(\pi)d\pi\} \, , \qquad \qquad
\phi(\pi) \equiv {1 \over \chi(\pi)-\pi} \, . \qquad \qquad  \
\label{f-pi}
\end{eqnarray}
\end{lemma}

The compressibility conditions H$_1$ and H$_2$ or $\bar{{\rm H}}_2$ have been stated, for general perfect fluids, in theorem \ref{theo-cc-nobar}. Now, in order to express H$_1$ for generic ideal gases,  where $\chi(\rho,p)=\chi(\pi)$, we have,
\begin{equation}\label{Derivchiresppi}
\chi_p' = \frac{\chi'(\pi)}{\rho} \, , \qquad \chi_{\rho}'=-\frac{\chi'(\pi)}{\rho}\pi \, , \qquad \chi\chi_p' + \chi_{\rho}'=\frac{\chi'(\pi)}{\rho}( \chi(\pi) - \pi) \, .
\end{equation}
Then, from the expression (\ref{cc-1-nonbar}) of the compressibility conditions H$_1$, we arrive to:
\begin{lemma}\label{lemma-ideal-cc1}
A generic ideal gas fulfills the relativistic compressibility condition ${\rm H}_1$ if, and only if, the square of the speed of sound $\chi=\chi(\pi)$ satisfies:
\begin{equation}\label{H1ideal}
{\rm H_1} :   \quad 0 < \chi < 1 \, , \quad   (1+\pi)(\chi-\pi) \chi'  + 2 \chi(1-\chi) > 0   \, .  \quad     \label{cc-1-ideal}
\end{equation}
\end{lemma}

On the other hand, H$_2$ or ${\rm \bar{H}}_2$ have been stated in theorem (\ref{theo-cc-nobar}) in terms of $r(\rho,p)$ and $s(\rho,p)$. For a generic ideal gas, where we have the expressions (\ref{r-s-ideal}) for these functions of state, we obtain:
$$
s_{\rho}' = - \frac{k \chi}{\rho( \chi- \pi)} \, ,  \qquad \frac{r_{\rho}'}{r} = \frac{\chi(\pi+1) - \pi}{\rho(\chi- \pi)(\pi+1)} \, .
$$
Consequently, if we substitute these expressions in (\ref{cc-2b-nonbar}), we have:
\begin{equation}\label{Idepi}
I(\rho,p) = \frac{\rho^2}{k \chi e(\pi)}[(2\pi +1)\chi - \pi] \quad .
\end{equation}
Taking into account that the first factor in (\ref{Idepi}) is positive, from expressions (\ref{Derivchiresppi}), (\ref{Idepi}) and theorem \ref{theo-cc-nobar}, it follows:
\begin{lemma}\label{lemma-ideal-cc2}
A generic ideal gas fulfills the relativistic compressibility condition ${\rm H}_2$ or $\bar{{\rm H}}_2$ if, and only if, the square of the speed of sound $\chi=\chi(\pi)$ satisfies:
\begin{equation}\label{H2barH2ideal}
{\rm H_2}: \quad \chi > \frac{\pi}{2\pi + 1}  \qquad \quad \mbox{ or } \qquad \quad   {\rm \bar{H}_2}: \quad  \chi < \frac{\pi}{2\pi + 1}  \, .
\end{equation}
\end{lemma}

Here we are interested in current generic ideal gases as defined by (\ref{gas-ideal}). Although we have seen that, from a formal point of view, equation (\ref{gas-ideal}) is compatible with compressibility condition $\bar{{\rm H}}_2$, from now on we will consider only the current compressibility conditions H$_1$ and H$_2$ for ideal gases suggested by the mentioned Lucquiaud's results. Thus, from the two lemmas above we obtain:
\begin{theorem}\label{propo-ideal-cc}
A generic ideal gas fulfills the relativistic compressibility conditions ${\rm H}_1$ and ${\rm H}_2$ if, and only if, the square of the speed of sound $\chi=\chi(\pi)$ satisfies:
\begin{equation}\label{H12ideal}
{\rm H_1} , {\rm H_2}:   \quad \frac{\pi}{2\pi + 1} < \chi < 1 \, , \quad   (1+\pi)(\chi-\pi) \chi'  + 2 \chi(1-\chi) > 0   \, .  \quad     
\end{equation}
\end{theorem}

This result allows us to characterize the perfect energy tensors that represent the evolution in l.t.e. of an ideal gas fulfilling the corresponding compressibility conditions. Now, for  similar reasons made before Theorems \ref{theo-cc-nonbar-chi} and \ref{theo-cc-nonbar-iso}, we analyze separately the isoenergetic and non isoenergetic evolutions. Then, if we take into account Lemma \ref{lemma-ideal} and Theorem \ref{propo-ideal-cc}, we arrive at the following result, that considers both cases together.

\begin{theorem} \label{theo-cc-ideal}
The necessary and sufficient condition for a non barotropic perfect energy tensor $T$ to represent the evolution in l.t.e. of a generic ideal gas submitted to the relativistic compressibility conditions {\rm H}$_1$ and {\rm H}$_2$ is:

\indent    i) if it is non isoenergetic, that the indicatrix function $\chi \equiv \dot{p}/\dot{\rho}$ depends on the sole hydrodynamic variable $\pi \equiv p/\rho$, $\chi=\chi(\pi) \not= \pi$, and it satisfies the  constraints {\rm (\ref{H12ideal})}.

\indent    ii) if it is isoenergetic, that it be isobaroenergetic, $\dot{\rho}=\dot{p}=0$. Then, the generic ideal gas thermodynamic quantities verifying these compressibility conditions are those for which the square of the speed of sound, $\chi = \chi(\pi)$, determined by these quantities, fulfills the constraints {\rm (\ref{H12ideal})}.
\end{theorem}

The above two theorems concern the compressibility condition ${\rm H}_2$ but, from lemma \ref{lemma-ideal-cc2}, two similar theorems can be stated for condition $\bar{{\rm H}}_2$ by changing the first inequalities in (\ref{H12ideal}) to:
\begin{equation}\label{H12barideal}
0 < \chi <  \frac{\pi}{2\pi + 1}   \, .  \quad     \label{cc-2bar-ideal}
\end{equation}

It is worth presenting some remarks about the results in this section.
\begin{itemize}
\item[{\bf i)}]
In the general inverse problem for a perfect energy tensor $T$, the inequalities {\rm (\ref{H1ideal})} guarantee H$_1$ if the square of the speed of sound is a function of the sole variable $\pi$, $\chi = \chi(\pi)$. 
Then, Lemma \ref{lemma-H1-H2} states that we can complete the thermodynamics with functions $r(\rho,p)$, $s(\rho,p)$ ensuring H$_2$ or $\rm\bar{H}_2$. But not {\em all} the resulting fluids will be generic ideal gases. The latter are the solution to the restricted inverse problem for generic ideal gases, they are given by Lemma \ref{lemma-ideal} and satisfy H$_2$ or $\rm\bar{H}_2$ if $\chi$ fulfills the corresponding inequality in {\rm(\ref{H2barH2ideal})}. Note that now, by means of the function $\chi$, H$_2$ or $\rm\bar{H}_2$ also restrict the hydrodynamic flow in this restricted inverse problem. 
\item[{\bf ii)}]
We have seen that the hydrodynamic variable $\pi = p/\rho$ plays a central role in the study of the properties of a generic ideal gas. In fact, for a general perfect fluid, the Pleba\'nski \cite{Plebanski} energy conditions state, $-p \leq \rho < p$, that is, they limit the range of physically admissible values of $\pi$ to $-1 \leq \pi \leq 1$. Under the reasonable assumption of positive matter density and temperature, the equation of state (\ref{gas-ideal}) implies a positive pressure and thus, the energy conditions for a generic ideal gas state that $0 < \pi < 1$. On the other hand, this is the range where the compressibility conditions H$_1$ is fulfilled for the thermodynamic fluids that do not fulfill ${\rm H_2}$ nor ${\rm \bar{H}_2}$. From (\ref{H2barH2ideal}), they are those whose square of the velocity of sound, say $\chi_*$, is $\chi_* = \pi/(2\pi+1)$. Thus, condition (\ref{H2barH2ideal}) states that {\em the graphic of $\chi(\pi)$ for the  generic ideal gases verifying the relativistic compressibility conditions ${\rm H}_1$ and ${\rm H_2}$ ({\em resp.} ${\rm \bar{H}_2}$) is over ({\em resp.} under) the graphic of $\chi_*(\pi)$ in the region $[0,1] \times (0,1)$.}
\item[{\bf iii)}]
Observe that Theorem \ref{theo-cc-ideal} only concerns the non barotropic case. Barotropic generic ideal gases have, necessarily  \cite{fluperLTE} \cite{coll}, the equation of state%
\footnote{We see here that in the present restricted inverse problem for generic ideal gases, the hypothesis of barotropicity completely determines the thermodynamics of the ideal gas.} %
$p=(\gamma-1) \rho$, and the compressibility conditions H$_1$ given by (\ref{cc-1-bar}) hold if, and only if, $1< \gamma<2$. But these ideal gases are physically meaningless in the common domain, because for them $\epsilon = c_v \Theta -1$ and the internal energy takes negative values at low temperatures.
\item[{\bf iv)}]
The expressions (\ref{H1ideal}) of the compressibility conditions  H$_1$  have been obtained from the corresponding general expressions (\ref{cc-1-nonbar}) under the hypothesis $\chi(\rho,p) = \chi(\pi)$. 
Thus (\ref{H1ideal}) also apply for a fluid with a speed of sound $c_s^2 = p/\rho$, that is $\chi = \pi$, and it is verified by this function for $\pi \in ]0,1[$. 
Nevertheless the expressions (\ref{H2barH2ideal}) of the compressibility conditions H$_2$ or $\rm\bar{H}_2$ do not apply in this case because expressions (\ref{r-s-ideal}) are only valid for ideal gases ($\chi \not= \pi$).
An open problem that will be considered elsewhere is to solve the inverse problem in this case, and to study the compressibility condition H$_2$ or $\rm\bar{H}_2$ for the associated thermodynamic schemes.
\item[{\bf v)}] As we mentioned in comment v) at the end of previous section, a non
 barotropic media failling to fulfill the expression (\ref{cc-1-nonbar}) of the compressibility conditions H$_1$ might have a barotropic evolution with a barotropic relation $p=\phi(\rho$) satisfying the expression (\ref{cc-1-bar}) of H$_1$ for barotropic fluids. As an example, let us consider a generic ideal gas with a square of the speed of sound $\chi( \pi)$ {\em which does not fulfill} the expression (\ref{H1ideal}) of the compressibility conditions H$_1$, let $e(\pi)$ be the specific energy given in (\ref{e-pi}), and let $\pi_0 \in]0,1[$ be a value of the hydrodynamic variable $\pi=p/\rho$. Then, from the results in \cite{fluperLTE} we obtain that an evolution at constant temperature $\Theta_0 = \frac{1}{k} \pi_0 e(\pi_0)$ leads to a barotropic evolution $p =\pi_0 \rho$ {\em which fulfills} the expression (\ref{cc-1-bar}) of H$_1$.
\end{itemize}
%

\section{Rainich approach to the ideal gas solutions}
\label{sec-Rainich}


The necessary and sufficient conditions for a metric $g$ to be a non-null Einstein-Maxwell solution were studied by Rainich \cite{Rainich}. We can distinguish the algebraic Rainich conditions, which impose that the Ricci tensor is of (non-null) electromagnetic type, and the differential ones, which guarantee that the electromagnetic field is a solution of the Maxwell equations.

A similar approach for the perfect fluid implies, in a first step, to obtain the algebraic conditions for a Ricci tensor to have a triple eigen-value, and a simple one with an associated time-like eigen-vector. These conditions and the additional ones assuring that the energy tensor fulfills the Pleba\'nsky \cite{Plebanski} energy conditions were obtained years ago \cite{bcm} \cite{Coll-Ferrando-termo}. A more recent version \cite{fs-ssst-Ricci} states:
\begin{lemma} \label{lemma-fluper-energy}

Consider the following scalar and tensor functions of the Ricci tensor $R$:
\begin{equation} \label{fluper-definitions}
 t \equiv \tr R  , \quad  N \equiv R - \frac14 t \, g  , \quad q \equiv - 2 \sqrt[3]{\frac{\tr N^3}{3}} , \quad Q \equiv N - \frac14 q \, g .
\end{equation}
A spacetime is a perfect fluid solution fulfilling the energy conditions if, and only if, the Ricci tensor $R$ satisfies:
\begin{equation} \label{fluper-conditions-A2}
Q^2 + q \, Q = 0  , \qquad  Q(x,x) > 0   ,  \qquad  q > 0   , \qquad  t+q > 0   ,
\end{equation}
where $x$ is any time-like vector.
\end{lemma}

\begin{lemma} \label{lemma-fluper-energy2}
When the conditions {\em(\ref{fluper-conditions-A2})} of Lemma \ref{lemma-fluper-energy} are verified,
the energy density $\rho$, the pressure $p$ and the unit velocity $u$ of the fluid are given by:
\begin{equation} \label{fluper-hydro}
 \rho = \frac14 (3 q + t)   , \qquad p = \frac14 (q - t)  , \qquad  u = \frac{Q(x)}{\sqrt{q \, Q(x,x)}}   .
\end{equation}
\end{lemma}
The extra conditions on the Ricci tensor for the fluid to be in l.t.e. were presented in \cite{Coll-Ferrando-termo}. They are obtained in terms of the Ricci invariants (\ref{fluper-definitions}) by writing their expression (\ref{h-lte}) with the values of $\rho$, $p$ and $u$ given by Lemma \ref{lemma-fluper-energy2} (see also  \cite{fs-ssst-Ricci}).

As we are concerned here by current generic gases, we must impose the energy conditions corresponding to a gas, $0 \leq \pi < 1$, $\pi=p/\rho$, as well as the l.t.e. condition, which states (see Lemma \ref{lemma-ideal}) $ \dif \chi \wedge \dif \pi = 0$, $\chi \not= \pi$. Moreover, the function $\chi(\pi)$ has to fulfill the compressibility conditions H$_1$ and H$_2$ for ideal gases given by Theorem \ref{propo-ideal-cc}.

When the perfect energy tensor is non isoenergetic, $\dot{\rho}\not=0$, the square of the speed of sound is given by the indicatrix function, $\chi= \chi \equiv \dot{p}/\dot{\rho}$. Then, taking into account Theorem \ref{theo-cc-ideal} and the expressions in Lemma \ref{lemma-fluper-energy}, we obtain the following.

\begin{theorem} \label{theo-Rainich-ideal}
Let us consider, in terms of the above scalar and tensor functions {\em (\ref{fluper-definitions})}  and {\em (\ref{fluper-hydro})} of the Ricci tensor $R$, the following ones:
\begin{equation} \label{fluper-definicions-ideal2}
 \pi  \equiv \frac{p}{\rho}   , \qquad \chi  \equiv \frac{Q( \dif p , \dif \rho)}{Q( \dif \rho , \dif \rho)}   , \qquad  \chi' \equiv  \frac{(\dif \chi,y)}{(\dif \pi,y)}   ,
\end{equation}
where $y$ is any transversal vector to $d\pi$: $(d \pi,y) \not= 0$.

A metric is a non isoenergetic perfect fluid solution modeling an ideal gas in l.t.e. that fulfills the compressibility conditions {\em H}$_1$ and {\em H}$_2$ if, and only if, the Ricci tensor $R$ satisfies:
\begin{eqnarray} \label{fluper-conditions-ideal}
Q^2 + q Q = 0  , \qquad  Q(x,x) > 0   ,  \qquad -t < q \leq t   , \\[2mm]
Q(\dif \rho) \not=0, \qquad  \dif \chi \wedge \dif \pi =0, \qquad \chi \not= \pi, \\[2mm]
\frac{\pi}{2\pi+1} < \chi < 1 \, , \qquad   (1+\pi)(\chi-\pi)  \chi'  + 2 \chi(1-\chi) > 0   \, .       \label{gasH1H2} 
\end{eqnarray}
where $x$ is any time-like vector. 
\end{theorem}

On the other hand, in the isoenergetic case, $\dot{\rho}=0$, Theorem \ref{theo-cc-ideal} states that the evolution is necessarily isobaric, $\dot{p}=0$. Moreover, any function $\chi(\pi)$ fulfilling (\ref{gasH1H2}) is compatible with this evolution. Consequently, we have:

\begin{theorem} \label{theo-Rainich-ideal-isoenergetic}
A metric is an isoenergetic perfect fluid solution modeling an ideal gas in l.t.e. that fulfills the compressibility conditions {\em H}$_1$ and {\em H}$_2$ if, and only if, the Ricci tensor $R$ satisfies {\em (\ref{fluper-conditions-ideal})} and
\begin{equation} \label{fluper-conditions-isoenergetic}
Q(\dif \rho) = 0  , \qquad  Q(\dif p) = 0   ,
\end{equation}
where $x$ is any time-like vector, and $Q$, $N$, $w$, $t$, $\rho$ and $p$ are given in {\em (\ref{fluper-definitions})}  and {\em (\ref{fluper-hydro})}. 
\end{theorem}
%


\section{Constraints on the ideal gas Stephani universes}
\label{sec-idealgas-Stephani}

The {\em Stephani universes} were obtained as the conformally flat perfect fluid solutions to Einstein
equations with nonzero expansion \cite{st}. Later, they were achieved \cite{ba} as the
conformally flat class of irrotational and shear-free perfect fluid space-times with nonzero expansion. They can also be characterized as the space-times verifying a weak cosmological principle without any hypothesis on the energy tensor \cite{bc}.

In \cite{C-F} we studied the Stephani universes that can be interpreted as an ideal gas in local thermal equilibrium. Now we summarize the main results in the following statements:
\ \\[1mm]
{\bf Thermodinamic Stephani universes} \cite{bc} The necessary and sufficient condition for a Stephani
universe to represent the evolution of a fluid in local thermal
equilibrium is that it admits a three-dimensional isometry group on two-dimensional orbits.
The metric of the thermodynamic Stephani universes may be written as:
\begin{equation}  \label{the-ste-uni}
d{\rm s}^2 = -\alpha^2 \dif t^2 + \Omega^2 \delta_{ij} \dif x^i \dif x^j \, ;
\end{equation}
\begin{equation} \label{eq:termetric}
\displaystyle \alpha \equiv R \partial_R \ln L \, , \quad
\Omega \equiv \frac{w}{2z} L\, , \quad   L \equiv \frac{R(t)}{1+ b(t) w} \, ,   \quad w
\equiv \frac{2z}{1 + {\varepsilon \over 4}{\rm r}^2}   \, ,
\end{equation}
$R(t)$ and $b(t)$ being two arbitrary functions of time. Its symmetry group is spherical, plane or pseudospherical
depending on $\varepsilon$ to be $1$, $0$ or $-1$ and the
Friedmann-Robertson-Walker limit occurs when $b=constant$. Furthermore, the energy density, the pressure, the expansion and the 3-space curvature are given by
\begin{equation}
\rho = \frac{3}{R^2} (\dot{R}^2 + \varepsilon - 4 b^2)   ,
\quad p= - \rho - {R \over 3} {\partial_R\rho \over \alpha}    , \quad  \theta = \frac{3\dot{R}}{R} \not=0     ,
\quad \kappa = \frac{1}{R^2} (\varepsilon - 4 b^2)     .
\label{tdp}
\end{equation}
\ \\[1mm]
{\bf Ideal gas Stephani universes} \cite{C-F} A thermodynamic Stephani universe {\rm (\ref{the-ste-uni})}\,{\rm
(\ref{eq:termetric})} represents an ideal gas if, and only if,
the metric function $b(R)$ and the energy density $\rho(R)$
satisfy the equations:
\begin{equation} \label{eq:eq12}
R a' = -a(c_{\scriptscriptstyle 1}a^2+c_{\scriptscriptstyle 2}a+1)
, \quad  b'' R = - c_{\scriptscriptstyle 1} a^2 b' , \quad a  =  a(R) \equiv  - \frac{R\, \partial_R\rho}{3 \rho} ,
\end{equation}
where the principal constants $c_{\scriptscriptstyle 1}$ and $c_{\scriptscriptstyle
2}$ are
arbitrary. Then, in terms of $b(R)$ and $\rho(R),$ the expansion factor
$R(t)$ satisfies the generalized Friedmann equation:
\begin{equation} \label{eq:Friedmann}
\rho(R) = \frac{3}{R^2}[ \dot{R}^2 + \varepsilon - 4 b^2(R)] .
\end{equation}
\ \\[1mm]
{\bf Ideal gas Stephani models} \cite{C-F} 
The indicatrix function $\chi=\chi(\pi)$ of the ideal gas Stephani universes takes the expression:
\begin{equation} \label{chi-stephani}
\chi(\pi) = \beta \pi^2 + \gamma \pi + \delta \, , \qquad \delta \equiv  \gamma - \beta - 2/3 ,
\end{equation}
where the constants $\beta$ and $\gamma$ determine the fundamental constants as:
\begin{equation}
c_1= 3 \beta, \qquad c_2 =3(\gamma - 2\beta -1)   .
\label{c1c2}
\end{equation}
Then, depending on the roots of the equation $\chi(\pi)=\pi$, we find five classes Cn of ideal gas Stephani universes:
\begin{itemize}
\item[] {\sc Class 1}: $c_1=c_2=0$.
\item[] {\sc Class 2}: $c_1=c_2\not=0$.
\item[] {\sc Class 3}: $\Delta \equiv c_2 - 4c_1=0$, $c_1\not=0$.
\item[] {\sc Class 4}: $\Delta \equiv c_2 - 4c_1>0$, $c_1\not=0$.
\item[] {\sc Class 5}: $\Delta \equiv c_2 - 4c_1<0$.
\end{itemize}
Moreover, for every class, we can obtain the thermodynamic quantities (\ref{e-t-ideal}) and (\ref{r-s-ideal}) by using the generating functions (\ref{e-pi}) and (\ref{f-pi}), and we obtain five different thermodynamics. 

On the other hand, the study of equations (\ref{eq:eq12}) leads to distinguish the singular models, ($a'(R)=0$), compatible with classes C2, C3, and C4, and regular models ($a'(R)\not=0$), compatible with the five classes Cn \cite{C-F}.

Now we will study the compatibility of the above ideal gas Stephani models with the compressibility conditions studied in Section \ref{sec-cc-idealgas}. It is worth remarking that a  {\em full} analysis of the physical behavior of the solutions implies, not only a survey of the Pleba\'nski energy conditions \cite{Plebanski}, $-p \leq \rho < p$, i.e.  $-1 < \pi \leq 1$, but also the knowledge of the solution to the generalized Friedmann equation (\ref{eq:Friedmann}), which allows us to obtain the coordinate dependence of $\rho$ and $p$ by (\ref{tdp}). But this is not our goal here.

Here we want to analyze when an ideal gas Stephani model fulfills the compressibility conditions H$_1$ and H$_2$ (i.e. verifies the corresponding  expressions (\ref{H12ideal})), {\em provided that} it fulfills the energy conditions for a gas:
\begin{equation}  \label{e-c-gas}
0 \leq \pi \leq 1 \, .
\end{equation}
We must look for the values of the principal constants $c_1$ and $c_2$ (or, equivalently, of the constants $\beta$ and $\gamma$) for which the indicatrix function (\ref{chi-stephani}) satisfies the expressions (\ref{H12ideal}) of the compressibility conditions H$_1$ and H$_2$ for a significant range of $\pi \in [0,1]$. 
We shall consider here separately the models with a physically reasonable behavior at low or high temperature. 


\subsection{Low temperature. Models with $\chi(0)=0$}

For an ideal gas, good performance at low temperatures means a vanishing speed of sound when pressure vanishes, that is $\chi(0)=0$. Then, from (\ref{chi-stephani}) and (\ref{c1c2}), we have:
\begin{equation} \label{chi-00}
\chi(\pi) = (\gamma-2/3) \pi^2 + \gamma \pi  \, ,
\end{equation}
\begin{equation} \label{c1c2-00}
c_1= 3 \gamma -2 \not=0, \qquad c_2 =1-3\gamma, \qquad \Delta  \equiv c_2^2-4c_1= 9(\gamma-1)^2   \, .
\end{equation}
Now we study the values of the parameter $\gamma$ for which the indicatrix function (\ref{chi-00}) fulfills the expressions (\ref{H12ideal}) of the compressibility conditions H$_1$ and H$_2$ in a neighborhood of $\pi=0$. Note that $\chi'(0)=\gamma$, and $\chi_*(0)=0$, $\chi_{*}'(0)=1$ for $\chi_*(\pi) = \pi/(2 \pi +1)$. 
%
%
 Then, if we impose the first inequality in (\ref{H12ideal}), $\chi_*(\pi) < \chi(\pi)$, we have, necessarily, $\gamma > 1$. On the other hand, a straightforward calculation shows that the second inequality in (\ref{H12ideal}), $\chi(\pi) <1$, holds in the interval $\pi \in [0, \pi_m[$, where
\begin{equation} \label{pi-max-00}
0<\pi_m = \frac{2}{\gamma + \sqrt{\gamma^2 +4(\gamma -2/3)}}  < 1 \, .
\end{equation}
Finally, under the already imposed constraints, the last inequality in (\ref{H12ideal}) is satisfied. Note that, for $\gamma>1$, one has $c_1\not=0$ and $\Delta >0$. Consequently, the ideal gas thermodynamics belong to class C4. Thus, we have:
\begin{proposition}  \label{propo-stephani-00}
The indicatrix function $\chi=\chi(\pi)$ of the ideal gas Stephani universes takes the expression {\em (\ref{chi-00})} when it verifies $\chi(0)=0$. When $\gamma>1$ the associated thermodynamics belong to class {\em C4} and fulfill the relativistic compressibility conditions {\em H$_1$} and {\em H$_2$}  for $\pi \in [0, \pi_m[$, where $\pi_m$ is given in {\em (\ref{pi-max-00})}.
\end{proposition}

Taking into account the results in \cite{C-F}  we can determine the generating functions (\ref{e-pi}) and (\ref{f-pi}) by using the indicatrix function (\ref{chi-00}). And, from them, we can obtain the associated ideal gas thermodynamics (\ref{e-t-ideal}) and (\ref{r-s-ideal}). These thermodynamics of class C4 are compatible with both, singular and regular models. The explicit expressions of these models can be found in \cite{C-F}, where the ideal gas Stephani models which approximate a classical ideal gas at first order in the temperature are studied. It is worth remarking that those models satisfy $\chi(0)=0$ and the parameter $\gamma$ coincides with the adiabatic index, which is submitted to $1<\gamma<2$. Nevertheless, the compressibility conditions also hold for $\gamma \geq 2$ in $\pi \in [0, \pi_m[$.


\subsection{High temperature. Models with $\chi(1/3)=1/3$}

For a relativistic Synge gas \cite{Synge} \cite{CFS-synge}, the limit $\Theta \rightarrow  \infty$ leads to a radiation fluid, $\rho=3p$, and a speed of sound $c_s = 1/\sqrt{3}$. Consequently, in order to have a good behavior at high temperatures we will impose to the indicatrix function $\chi$ the constraint $\chi(1/3)=1/3$. Then, from (\ref{chi-stephani}) and (\ref{c1c2}), and introducing the parameter $\lambda= 7-8 \gamma $ , we have:
\begin{equation} \label{chi-33}
\chi(\pi) = \frac{1}{16}[3(1-\lambda) \pi^2 +  2(7 - \lambda) \pi + (\lambda+ \frac13)] \, ,
\end{equation}
\begin{equation} \label{c1c2-00}
c_1= \frac{9}{16}(1- \lambda), \qquad c_2 =\frac34 (\lambda -2), \qquad \Delta  \equiv c_2^2-4c_1= \frac{9}{16} \lambda^2   \, .
\end{equation}
We can see that, for any value of the parameter $\lambda$, the indicatrix function (\ref{chi-33}) fulfills the expressions (\ref{H12ideal}) of the compressibility conditions H$_1$ and H$_2$ in $\pi=1/3$, and, therefore, in its neighborhood. Moreover, depending on the values of $\lambda$, the ideal gas thermodynamics belong to classes C2, C3 or C4. Thus, we have:
\begin{proposition}  \label{propo-stephani-33}
The indicatrix function $\chi=\chi(\pi)$ of the ideal gas Stephani universes takes the expression {\em (\ref{chi-33})} when it verifies $\chi(1/3)=1/3$. For any value of the parameter $\lambda$ the associated thermodynamics fulfill the relativistic compressibility conditions {\em H$_1$} and {\em H$_2$} in a neighborhood of $\pi=1/3$, and they belong to class {\em C2} when $\lambda=1$, to class {\em C3} when $\lambda= 0$, and to class {\em C4} otherwise.
\end{proposition}

Taking into account the results in \cite{C-F} and by using the indicatrix function (\ref{chi-33}) we can determine the generating functions (\ref{e-pi}) and (\ref{f-pi}) and, from them, we can obtain the associated ideal gas thermodynamics (\ref{e-t-ideal}) and (\ref{r-s-ideal}). These thermodynamics of class C2, C3 and C4 are compatible with both singulars and regular models. Here, we do not give the explicit expressions of all these models. We restrict ourselves to the cases which also are near to the Synge gas at the following order in temperature.

The derivative of the indicatrix function $\chi=\chi(\pi)$ of a Synge gas satisfies \cite{Synge} \cite{CFS-synge} $\chi'(1/3) = 1/2$. From (\ref{chi-33}) we have:
\begin{equation} \label{chiprima-33}
\chi'(\pi) = \frac{1}{8}[3(1-\lambda) \pi +  (7 - \lambda)]  \, , \qquad   \chi''(\pi) = \frac38(1-\lambda) \, .
\end{equation}
If we impose $\chi'(1/3)$ to be in the neighborhood $]1/4, 3/4[$ of $1/2$, then the parameter $\lambda$ satisfies $1 < \lambda < 3$, and $c_1 \not=0$ and $\Delta >0$. Consequently the thermodynamics belong to class C4.

Before determining the domain where the indicatrix function (\ref{chi-33}) fulfills 
the expressions (\ref{H12ideal}) of the compressibility conditions H$_1$ and H$_2$, we will obtain the generating functions (\ref{e-pi}) and (\ref{f-pi}) by using the indicatrix function (\ref{chi-33}). A straightforward calculation leads to:
\begin{equation} \label{ef-33}
e(\pi) = e_0  \frac{|3(\lambda-1) \pi + 3 \lambda +1|^{3 + \frac{1}{\lambda}}}{|1-3 \pi|^{\frac{1}{\lambda}} (\pi + 1)^3}  \, ,
\qquad    f(\pi) = f_0 \left| \frac{3(\lambda-1) \pi + 3 \lambda +1}{1-3 \pi } \right|^{\frac{4}{\lambda}}  .
\end{equation}
Note that both generating functions diverge in $\pi=1/3$, and from (\ref{e-t-ideal}), the temperature also diverges as occurs in the relativistic Synge gas, where $\pi \leq 1/3$. If we look for a model which approximates a Synge gas we must restrict ourselves to $\pi \in [0, 1/3[$. Now we shall see when the indicatrix function (\ref{chi-33}) fulfills the expressions (\ref{H12ideal}) of the compressibility conditions H$_1$ and H$_2$ in this domain.

We have $\chi'(\pi) > 0$ in $[0, 1/3[$, and $\chi(0) = (\lambda+1/3)/16>0$, $\chi(1/3) = 1/3$. Thus $0<\chi(\pi) \leq 1/3 <1$. The indicatrix function depends on the parameter $\lambda$ and we can consider it as a function of two variables $\chi (\pi, \lambda)$. We have $\frac{\partial \chi}{\partial \lambda} > 0$ in the considered domain and, consequently, $\chi(\pi, \lambda) \geq \chi(\pi,1)$. We can easily check that $\chi(\pi,1) \geq \chi_*(\pi) = \frac{\pi}{2 \pi +1}$. Therefore, the first compressibility condition in (\ref{H12ideal}) holds. 

On the other hand, if $\pi \in [0,1/3[$, we have $\chi(\pi) \leq \chi(\pi, 1) \leq \pi$, $\chi'(\pi)>0$, and $0 < \chi(\pi) < 1$, and consequently the second compressibility condition in (\ref{H12ideal}) also holds.

\begin{proposition}  \label{propo-stephani-332}
The indicatrix function $\chi=\chi(\pi)$ of the ideal gas Stephani universes takes the expression {\em (\ref{chi-33})} when it is submitted to $\chi(1/3)=1/3$. If additionally the parameter $\lambda \in ]1,3[$ (or equivalently, $\chi'(1/3) \in ]1/4, 3/4[$, i.e. it is closed to a Synge gas), then the associated thermodynamics fulfill the relativistic compressibility conditions H$_1$ and H$_2$ in $\pi=]0,1/3[$, and they belong to class {\em C4}.
\end{proposition}
The ideal gas models of class C4 in proposition above demostrate good physical behavior if the hydrodynamic variable $\pi = p/\rho$ ranges in $[0,1/3[$, and they approach a Synge gas at high temperature, that is in a neighborhood of $1/3$. Nevertheless, at low temperatures they differ from the Synge gas because the speed of sound does not approach zero when the temperature reaches zero.

From the generating functions (\ref{ef-33}) we can obtain the associated ideal gas thermodynamics (\ref{e-t-ideal}) and (\ref{r-s-ideal}). These thermodynamics of class C4 are compatible with both singulars and regular models. The explicit expressions of these models can be easily obtained from the results in \cite{C-F}. By way of example, here we make explicit the generalized Friedmann equations for one of the singular models compatible with the case $\lambda=2$ (the best fit for a Synge gas):
\ \\[1mm]
{\bf Generating functions}
\begin{equation} \label{ef-33-2}
e(\pi) = e_0  \frac{(3 \pi + 7)^{\frac{7}{2}}}{(1-3 \pi)^{\frac{1}{2}} (\pi + 1)^3}  \, ,
\qquad    f(\pi) = f_0 \left( \frac{3 \pi + 7}{1-3 \pi } \right)^2 \, .
\end{equation}
\ \\[1mm]
{\bf Generalized Friedmann equations}: 
\be
\rho(R) = \frac{3}{R^2}[ \dot{R}^2 + \varepsilon - 4 b^2(R)] \, ,
\ee
where
\begin{equation} \label{Friedmann-2}
\rho (R) = \rho_0  \left( \frac{R_0}{R} \right)^4  ,
\qquad   b (R) = b_1 + b_2 \left( \frac{R}{R_0} \right)^2  .
\end{equation}
%


\section{Ending comments}
\label{sec-remarks}

As we have seen in Section \ref{sec-t-flow}, a thermodynamic perfect fluid in l.t.e. may be considered as a conservative and deterministic hydrodynamic fluid endowed with subsidiary thermodynamic quantities. We know that causal and energy conditions  directly constraint hydrodynamic variables and, consequently, they restrict the physically admissible energy tensors and, then, the  gravitational fields that may be related to them. But whether or not the compressibility conditions, involving thermodynamic variables, also restrict the physically admissible energy tensors, remained an open problem, which we have solved in Section \ref{sec-cc} (Theorems \ref{theo-cc-bar}, \ref{theo-cc-bar-b}, \ref{theo-cc-nobar}, \ref{theo-cc-nonbar-chi} and \ref{theo-cc-nonbar-iso}). We have seen that, at most, only H$_1$ generically restricts them, and we have shown the form of the corresponding constraints. These results correspond to the conceptual point of view referred to in the Introduction.

An important task in Relativity is the study of the physical meaning of the formal solutions $T = \{u, \rho, p\}$ to the conservative perfect fluid equations $\nabla \cdot T =0$, whether or not they are coupled to the gravitational field by Einstein equations. At present, a wide family of such solutions is known without specific physical meanings.
Our hydrodynamical approach (see also \cite{fluperLTE}) provides a tool to analyze the physical reality for this family of solutions and to solve the inverse problem for obtaining the specific thermodynamic interpretation.
We have shown that the compressibility conditions H$_1$ impose constraints on the hydrodynamic quantities $\{u, \rho, p\}$, while the compressibility condition H$_2$ only imposes constraints on the thermodynamic subsidiary quantities.

We have applied these results to the particular case of generic ideal gases. In this case, because of the very definition of generic ideal gases, where hydrodynamic and thermodynamic quantities are involved, not only H$_1$ but also H$_2$ constrain the hydrodynamic quantities (Theorems \ref{propo-ideal-cc}  and \ref{theo-cc-ideal}). 

Is is worth remarking that the above approach offers an IDEAL (Intrinsic, Deductive, Explicit and ALgorithmic) characterization of the perfect energy tensors that model the evolution of a physically realistic perfect fluid. In \cite{fluperLTE} we outlined the interest and the applications of this kind of approach.

A Rainich-like theory of a physical medium is the {\em complete geometrization} of the physical relations characterizing it, that is to say, the formulation of these relations (equations and inequalities) in terms of concomitants of the sole space-time metric $g$.
We have here formulated the Rainich approach to generic ideal gases submitted to the compressibility conditions  H$_1$ and H$_2$ (Theorems \ref{theo-Rainich-ideal} and \ref{theo-Rainich-ideal-isoenergetic}).

Another application is the determination of the metrics of a given family of perfect fluid solutions that represent the evolution of a specific set of fluids. The physical requirements studied here for generic ideal gases have been imposed on the ideal gas Stephani universes already obtained elsewhere \cite{C-F} and  physically realistic models have been selected (Sect. \ref{sec-idealgas-Stephani}). A similar study has been performed elsewhere for the Szskeres-Szafron solutions of class II \cite{CFS-PSS}.

These applications correspond to the practical point of view referred to in the Introduction.


\begin{acknowledgements}
This work has been supported by the Spanish ``Ministerio de
Econom\'{\i}a y Competitividad", MICINN-FEDER project FIS2015-64552-P.
\end{acknowledgements}

\end{document}